# Practically useful form of Kim model from hysteresis loop of a superconductor

Ratan Lal

Superconductivity Division, National Physical Laboratory, Council of Scientific and Industrial Research, New Delhi-110012, INDIA

**Abstract**

Considering a theoretical hysteresis loop within the Kim model for the Kim constant $H_0$=0 it has been found that the expression of $H_0$ obtained in Lal [R. Lal, Physica C **470** (2010) 281] provides a nonzero value $0.3424H_p$ for this constant. ($H_p$ is full penetration field.) This different value of the Kim constant for the same hysteresis loop has been made a base for a different version of the Kim model, the hysteretic Kim model, such that the value zero of $H_0$ corresponds to the original Kim model and the value $0.3424H_p$, denoted by a different notation $H_{0,hys}$, corresponds to the hysteretic Kim model. The two versions of the Kim model are interrelated so that $H_{0,hys}$ is a function of $H_0$. An empirical relation, $H_{0,hys} = 0.3424H_p \exp(1.4\sqrt{H_0/H_p})$, has been worked out on the basis of the theoretical $H_0$>0 hysteresis loops. The hysteretic Kim model has been cast in a practically useful form by obtaining an expression of $H_p$ in terms of the hysteretic magnetization. The importance of the hysteretic Kim model has been illustrated by applying it to the $YBa_2Cu_3O_7$, $Bi_2Sr_2CaCu_2O_{8+\delta}$ and $Ba_{0.72}K_{0.28}Fe_2As_2$ superconductors by taking hysteresis loops of these systems from literature. It has been found that in these superconductors the dependence of $H_p$ on the sample width is mainly like $\sqrt{2a}$, and not like $2a$ (Bean model). ($2a$ is the sample width.) The empirical relation of $H_{0,hys}$ and $H_0$ has been found to provide a reasonably good understanding of the intergranular matrix of the $YBa_2Cu_3O_{7-\delta}$ superconductor, where Kim model has not been found successful earlier.

Telephone No.: (+) 91-11-45608360

E-mail address: rlal_npl_3543@yahoo.in

## 1. Introduction

In 1962 Bean [1] presented a critical state model for describing the magnetization of a hard superconductor by assuming that the critical current density is independent of the magnetic field. In the same year Kim, Hempstead and Strnad [2] extended the bean model by incorporating magnetic field dependence in the critical current density. Unfortunately the Kim model [2] did not turn out as practically useful as the Bean model. For example, while analyzing their full penetration field data of the $Bi_2Sr_2CaCu_2O_{8+\delta}$ superconductor, Wang et al [3] used the Bean model [1,4] although, as shown below, the Kim model provides a much better understanding of their data. In fact, the studies based on the Kim model remained limited mainly to the theoretical type of work in the areas like the low-field AC susceptibility [5-8] and fishtail effect [9].

There are two main reasons for the less practical usefulness of the Kim model. The first is that the two constants $k$ and $H_0$, which are involved in this model, have only one equation between them, namely [10]

$$J_{c,Kim}(H) = k/(H_0 + H), \qquad (1)$$

where $J_{c,Kim}(H)$ is the critical current density in the Kim model. But a unique solution of $k$ and $H_0$ requires two equations between them.

The second main reason lies in the fact that for $H_0=0$ and $H=0$ Eq. (1) leads to an infinite value, which is unphysical. The case of $H_0=0$ arises, for example, in the full penetration field data of Wang et al [3] (see below). In fact, the critical current density has an upper bound given by $\eta J_D$, where $\eta$ is the strength of the pinning force density [11] and $J_D$ is the depairing current density [12]. In order that the critical current density remains in its physical range, the Kim constant $H_0$ should be larger than $k/\eta J_D$.

In this article we show that it is possible to get rid of the above two problems in favor of a consistent and practically useful version of the Kim model by using the hysteretic magnetization. Details of our method are presented in Sec. 2. A brief account of the method is as follows. We start from the theoretical hysteresis loop for $H_0=0$, and then apply the expression of this constant from Ref. [10]. In this reference Kim model (Eq. 1) is used in terms of hysteretic magnetization for low $H$ so that the critical current density is always finite. Below we shall see that the critical current density assumes a physically reasonable value even for $H_0=0$. We find that the expression of $H_0$ in Ref. [10] provides a nonzero value $0.3424H_p$ for $H_0$, where $H_p$ is the full penetration field. Since we took the considered hysteresis loop corresponding to $H_0=0$, we treat this nonzero value of the Kim constant to correspond to a different version of the Kim model, namely to the hysteretic Kim model. Accordingly, we use a different notation, $H_{0,hys}$, for $H_0$ in the hysteretic Kim model. Thus, when $H_0=0$ in the original Kim model, then $H_{0,hys}=0.3424H_p$ in the hysteretic Kim model. We treat $H_{0,hys}$ as a function of $H_0$ for all values of $H_0$ such that $H_{0,hys}(H_0=0)=0.3424H_p$ and $H_{0,hys} \to H_0$ for $H_0 \to \infty$. For a certain range of $H_0$ we have found an empirical relation (c. f. Eq. 12 below) between $H_{0,hys}$ and $H_0$.

The other Kim constant $k$ has also been denoted by a different notation, $k_{hys}$, in the hysteretic Kim model. According to Ref. [10] the Kim constants $H_{0,hys}$ and $k_{hys}$ require $H_p$ for their determination. So, we have obtained a practically useful expression (c. f. Eq. 20 below) for $H_p$ in terms of the hysteretic magnetization.

In Sec. 3 we apply the hysteretic Kim model to the $YBa_2Cu_3O_7$, $Bi_2Sr_2CaCu_2O_{8+\delta}$ and $Ba_{0.72}K_{0.28}Fe_2As_2$ superconductors, and clarify the importance and practical usefulness of the present method. One of the main results of our study is that $H_p$ varies mainly like $\sqrt{2a}$ in the considered superconductors, and therefore the Bean result $H_p=J_{c,Bean}a$ [1,4] is too bad for these superconductors. Here $2a$ is the sample width, and $J_{c,Bean}$ is the magnetic field independent critical current density of the Bean model.

**2. Theory**

2.1. Hysteretic solution for $H_0 < k/\eta J_D$

Let us first clarify that $H_0 = 0$ is a realistic case. For this purpose we consider the full penetration field data of Wang et al [3]. These authors have measured the full penetration field of the $Bi_2Sr_2CaCu_2O_{8+\delta}$ superconductor of various sizes at different temperatures. For specificity, we consider $H_p$ measured by these authors for the 90×90 μm² and 50×50 μm² samples at the temperature $T$=8 K. From Fig. 6 of these authors we find that the values of $H_p$ are $\mu_0 H_{p1}$=0.30 T and $\mu_0 H_{p2}$=0.22 T for the 90×90 μm² and 50×50 μm² samples respectively. (Here $\mu_0=4\pi\times10^{-7}$ H/m is the free space permittivity.) The corresponding sample widths are obvious as $2a_1$=90 μm and $2a_2$=50 μm. On the basis of these values we find that $\mu_0 H_{p1}/\mu_0 H_{p2}$=0.30 /0.22=1.36, $2a_1/2a_2$=90/50=1.8, and $\sqrt{2a_1/2a_2} = \sqrt{90/50} = 1.34$. A comparison of these ratios shows that in the considered $Bi_2Sr_2CaCu_2O_{8+\delta}$ superconductor [3] $H_p$ follows the $\sqrt{2a}$ dependence on the sample width more closely, and that of the Bean model, namely $H_p=J_{c,Bean}a$ [1,4]. On the other hand, Wang et al [3] have argued that the Bean result is approximately followed in their Bi$_2$Sr$_2$CaCu$_2$O$_{8+\delta}$ superconductor. The point is that these authors have not tried to analyze their data in the Kim model, perhaps due to the problems mentioned in the previous section. Below we shall see that in the Kim model it is quite possible to obtain a $\sqrt{2a}$ sample-width-dependence of $H_p$ [6].

According to the Kim model the full penetration field $H_p$ is given by [13]

$$H_p = \sqrt{H_0^2 + 2ka} - H_0 \qquad (2)$$

for a sample of cross section 2a×2b ($a \leq b$).

Eq. (2) results in a $\sqrt{2a}$ sample-width-dependence of $H_p$ when $H_0=0$. In this sense the Bi$_2$Sr$_2$CaCu$_2$O$_{8+\delta}$ superconductor of Wang et al [3] appears to follow the Kim model with $H_0=0$. This is an example that $H_0=0$ in particular, and $H_0 < k/\eta J_D$ in general, are real cases. But we have mentioned above that $H_0 < k/\eta J_D$ corresponds to an unphysical critical current density. Therefore, there is a need for physically reasonable theory of hard superconductors for $H_0 < k/\eta J_D$.

We proceed towards a possible solution of this problem first for $H_0 = 0$. The case of $H_0 > 0$ will be considered thereafter. We start by considering the $H_0 = 0$ hysteresis loop in the Kim model. Let $M^+(H)$ and $M^-(H)$ denote respectively the positive and negative parts of the magnetization of this hysteresis loop. An expression for $M^-(H)$ has been obtained in Ref. [10] for $H_0=0$ and $0 \leq H \leq H_p$. Following the same method we obtain an expression of $M^+(H)$. The resulting $M^+(H)$ and Eq. (4) of Ref. [10] may be written as

$$M^+(h) = (H_p/15)\left[-15h - 20h^3 - 8h^5 + 8(1+h^2)^{\frac{5}{2}}\right] \qquad (3)$$

and

$$M^-(h) = (H_p/15)\left[-15h + 20h^3 - 8h^5 - 8(1-h^2)^{\frac{5}{2}}\right] \tag{4}$$

for $H_0 = 0$ and $0 \leq h \leq 1$ with $h = H/H_p$ as the reduced magnetic field.

According to Eq. (18) of Ref. [10] the Kim constant $H_0$ is expressed in terms of the vertical *M-H* loop width

$$W_p = M^+(H_p) - M^-(H_p), \tag{5}$$

and the non-vertical *M-H* loop width

$$W_{max} = M^+(-H_{min}) - M^-(H_{min}) = 2M^+(-H_{min}) = -2M^-(H_{min}). \tag{6}$$

Here $H_{min}$ is the minimum value of $M^-(H)$, which lies in the fourth quadrant of the *M-H* loop as is clear from Fig. 1. Notice that the various forms of $W_{max}$ in Eq. (6) arise due to the $M^+(H) = -M^-(-H)$ symmetry in the Kim model.

From Eqs. (3) and (4) we find that $W_{max}=1.3735H_p$ and $W_p=0.3503H_p$. Putting these values in Eq. (18) of Ref. [10] we find that $H_0= 0.3424H_p$. But we took the hysteresis loop of Eqs. (3) and (4) corresponding to $H_0= 0$. This means that we have got two values of $H_0$ for the same hysteresis loop. In fact these two values of $H_0$ correspond to two different situations. The first situation corresponds to the critical current density as given by the Kim model according to Eq. (1) without an involvement of the hysteresis loop [13]. In this case the critical current density may be infinitely large also, for example when $H_0$ is near zero. On the other hand, the second situation is governed by the physically reasonable critical current density in accordance with the hysteresis loop in a way described in Ref. [10]. In view of this difference in the two situations of the values of $H_0$, we consider two different versions of the Kim model. The first one is the original Kim model which corresponds to $H_0= 0$. In the second version we use a different notation for $H_0$ namely $H_{0,hys}$ and call this version as hysteretic Kim model to emphasize that in this model the critical current density is governed by the hysteresis loop. We emphasize that these two models are not independent because $H_{0,hys}$ depends on $H_0$ in that when $H_0= 0$ (in the original Kim model), then

$$H_{0,hys}(H_0= 0)=0.3424H_p \qquad (7)$$

in the hysteretic Kim model. For other values of $H_0$ there will be different values of $H_{0,hys}$ (see Eq. 12 below). For a given hysteresis loop (i.e., for a given $H_0$) $H_{0,hys}$ is given by Eq. (18) of Ref. [10] which is written as

$$H_{0,hys} = H_p W_p/(W_{max} - W_p). \qquad (8)$$

It may be noted that according to Eq. (6) $W_{max}$ is connected with the peak of $M^+(H)$ which lies at $H = -H_{min}$, or equivalently with the dip (lying at $H=H_{min}$) of $M^-(H)$. In this sense $W_{max}$ will increase with increasing peak height of $M^+(H)$. If there is no peak structure in $M^+(H)$ then $W_{max} = W_p$, implying $H_{0,hys} \to \infty$. This limiting value of $H_{0,hys}$ corresponds to the Bean model [13]. Since the Bean model indeed does not have a peak structure in $M^+(H)$ (c. f., e. g. Fig. 6a of Ref. 13) it may be said that Eq. (8) is a consistent expression of $H_{0,hys}$.

For a complete specification of the hysteretic Kim model we need different specifications of $k$ and $J_{c,Kim}$ also. In the hysteretic Kim model we denote $k$ and $J_{c,Kim}$ respectively by $k_{hys}$ and $J_{c,hys}$. According to Eq. (17) of Ref. [10] $k_{hys}$ will given by

$$k_{hys} = GW_{max}H_{0,hys}/a \qquad (9)$$

where

$$G = 3b/(3b - a) \qquad (a \leq b) \qquad (10)$$

is a geometric factor.

According to Ref. [10], the expression of the critical current density $J_{c,hys}$ has the same mathematical form as Eq. (1) so that we may write

$$J_{c,hys}(H) = k_{hys}/(H_{0.hys} + H) \qquad (11)$$

We now turn to a relation of $H_{0,hys}$ and $H_0$. According to Eq. (8) the Kim constant $H_0$ is involved implicitly in $W_{max}$ and $W_p$. Thus Eq. (8) is an implicit relation between $H_{0,hys}$ and $H_0$. In order to obtain an explicit relation between $H_{0,hys}$ and $H_0$ for $H_0 > 0$ we proceed as follows. From the hysteresis loops of Fig. 6 of Chen and Goldfarb [13] we find that $H_{0,hys}/H_p = 0.40, 0.53, 0.88$ and 2.86 for $H_0/H_p = 0.001, 0.11, 0.46$ and 2.41 respectively. These values and Eq. (7) are represented by the empirical relation

$$H_{0,hys} = 0.3424 H_p \exp\left(1.4\sqrt{H_0/H_p}\right) \qquad (H_0/H_p \leq 2.41) \qquad (12)$$

with about a 95% accuracy.

Eq. (12) provides the sought-for set of the values of $H_{0,hys}$ for $H_0 < k/\eta J_D$. Here it may be noted that the values of $H_0$ given by $H_0 < k/\eta J_D$ lie near $H_0 = 0$ so that the restriction of $H_0/H_p \leq 2.41$ for which Eq. (12) is obtained is not a problem.

It may be realized that Eq. (12) serves as a bridge between the Kim model and the hysteretic Kim model. This is because if one knows the values of $H_0$ then one can obtain values of $H_{0,hys}$, and vice versa. Below we shall use this bridge equation in both ways. At present let us apply Eq.

(12) to the $Bi_2Sr_2CaCu_2O_{8+\delta}$ superconductor of Wang et al [3], which we have found above to follow the Kim model with $H_0=0$. For specificity we consider the 90×90 μm² sample at the temperature $T$=8 K for which the value of $H_p$ is $\mu_0H_p$=0.30 T (see above). According to the bridge equation, Eq. (12), $H_0$=0 and $\mu_0H_p$=0.30 T leads to $\mu_0H_{0,hys}$=0.103 T. Corresponding to this value of $H_{0,hys}$ Eq. (9) and (10) (with $b=a$) give $k_{hys}$ = 1.08×10$^{15}$ A²/m³. These values of $H_{0,hys}$ and $k_{hys}$ lead Eq. (11) to $J_{c,hys}$ ($H$=0) = $k_{hys}$/ $H_{0,hys}$ = 1.31×10$^{10}$ A/m². On the other hand, $J_{c,Kim}$ ($H$=0) = ∞ since $H_0$=0. Thus we have not only got a finite value of the critical current density, we have also got a value of $J_{c,hys}$ ($H$=0) which is almost the same as found by Wang et al [3]. In fact, this is a typical value of the critical current density for the $Bi_2Sr_2CaCu_2O_{8+\delta}$ superconductor [14]. Thus, while the Kim model gives an unphysical large value of the critical current density for the $Bi_2Sr_2CaCu_2O_{8+\delta}$ superconductor, the hysteretic Kim model provides a physically reasonable value. This shows the importance of the bridge equation.

2.2. Critical current density

The first introduction of the critical current density was given by Bean [1]. He assumed the critical current density to be a constant given by [1, 4]

$$J_{c,Bean} = H_p/a = G\Delta M_0/a \qquad (13a)$$

Here $\Delta M_0$ is the vertical width of the hysteresis loop. It is independent of $H$ because according to Bean model both $M^+(H)$ and $M^-(H)$ are independent of $H$. When the magnetic field dependence of $M^+(H)$ and $M^-(H)$ is considered phenomenologically in Eq. (13a) the critical current density acquires a dependence on the magnetic field, which we may write as

$$J_{c,BM}(H) = G\Delta M(H)/a \tag{13b}$$

Here

$$\Delta M(H) = M^+(H) - M^+(H) \tag{14}$$

is the vertical width of the hysteresis loop at the magnetic field *H*.

Eq. (13b), known as Bean model critical current density, is widely used in literature for calculating the critical current density of a superconductor [10, 14-17]. However, in Ref. [10] we have argued that the critical current density given by the Bean model, Eq. (13b), is inconsistent for low *H* in many cases, and that up to $H=H_p$ a consistent form of the critical current density is given by Eq. (11). Here it must be noted that the critical current density in the original Kim model differs from that in the Bean model below $H=3H_p/2$ [13]. But in the hysteretic Kim model the critical current density will, deviating from the original Kim model tend towards the Bean model value. In order to see how it happens we proceed as follows.

From Eqs. (1) and (2) we find that

$$H_p = 2a\left[\frac{1}{J_{c,Kim}(0)} + \frac{1}{J_{c,Kim}(H_p)}\right]^{-1} \tag{15}$$

We also need an analog of this equation in the hysteretic Kim model. For this purpose first of all we notice that the analog of Eq. (2) in the hysteretic Kim model will be

$$H_p = \sqrt{H_{0,hys}^2 + 2k_{hys}a} - H_{0,hys}. \tag{16}$$

From this we obtain

$$H_p = 2a\left[\frac{1}{J_{c,hys}(0)} + \frac{1}{J_{c,hys}(H_p)}\right]^{-1} \tag{17}$$

We emphasize that Eq. (15) or Eq. (17) does not mean that $H_p \propto 2a$ as in the Bean model. This is because $H_p$ is also involved in the right-hand-side of these equations through $J_{c,hys}(H_p)$ or through $J_{c,Kim}(H_p)$. In fact the actual dependence of $H_p$ on the sample width is given by Eq. (2) or by Eq. (16). For extracting information from Eqs. (15) and (17), we first of all write expressions of $J_{c,Kim}(0)$ and $J_{c,hys}(0)$ by using Eqs. (1), (2), (11) and (16). These expressions are

$$J_{c,Kim}(0) = \frac{k}{H_0} = \frac{H_p}{a} + \frac{H_p^2}{2aH_0} \tag{18}$$

and

$$J_{c,hys}(0) = \frac{k}{H_{0,hys,}} = \frac{H_p}{a} + \frac{H_p^2}{2aH_{0,hys}} \tag{19}$$

Since according to Eq. (12) $H_{0,hys} > H_0$, we find from Eqs. (18) and (19) that $J_{c,hys}(0) < J_{c,Kim}(0)$. Then, Eqs. (15) and (17) provide $J_{c,hys}(H_p) > J_{c,Kim}(H_p)$. But according to Fig. 6 of Chen and Goldfarb [13] $J_{c,BM}(H_p) > J_{c,Kim}(H_p)$ for all $H_0$. This means that $J_{c,hys}(H)$ turns out to be closer to $J_{c,BM}(H)$ near $H=H_p$. This means that the range of agreement of $J_{c,hys}(H)$ and $J_{c,BM}(H)$ will be up to a magnetic field lower than $3H_p/2$. Although we are not sure about the exactness of the range of the agreement of $J_{c,hys}(H)$ and $J_{c,BM}(H)$, for specificity we assume that it is up to $H = H_p$ from the higher magnetic field side. In this sense a more reasonable form of the critical current density will be $J_{c,hys}(H)$ up to $H = H_p$, and $J_{c,BM}(H)$ above $H_p$.

2.3. Full penetration field $H_p$

In order to find out $H_{0,hys}$ (Eq. 8) and $k_{hys}$ (Eq. 9) we need a practical expression of the full penetration field because it is involved in both of these constants. For obtaining the required expression we put expressions of $H_{0,hys}$ and $k_{hys}$ respectively from equations (8) and (9) in Eq. (16). Then making algebraic simplifications we find

$$H_p = 2GW_{max}W_p/(W_{max} + W_p). \qquad (20)$$

Here $G$ is given by Eq. (10).

Let us evaluate Eq. (20) in the Bean limit where $M^+(H)$ and $M^-(H)$ are independent of $H$ so that $W_{max} = W_p = \Delta M_0$. Using these values in Eq. (20) we obtain Eq. (13a). This shows that the expression of the full penetration field given by Eq. (20) is a consistent expression.

Since the right-hand side of Eq. (20) also involves $H_p$ through $W_p$, this equation can be solved only numerically. When the hysteresis loop has a feature similar to that of the Kim model (c. f., Fig. 6 of Ref. [13]), the numerical solution is expected to be an easy process. Let us illustrate this for the hysteresis loop of Fig. 1 which has features of the Kim model. We use a multistep numerical method by starting with the initial value $H_p = 0$ by assuming that $b = a$. First of all we find the non-vertical loop width $W_{max}$ (Eq. 6) from Fig. 1, which is $W_{max} = 2 \times 3.70 \times 10^5$ A/m = $7.40 \times 10^5$ A/m. The value of $W_p$ for $H_p = 0$ is found from Fig. 1 to be $W_p = 5.40 \times 10^5$ A/m. Substituting these values in Eq. (20) we get $\mu_0 H_p = 1.17$ T. The second step starts with $\mu_0 H_p = 1.17$ T and $W_p(H = H_p) = 2.77 \times 10^5$ A/m. With these values, and the fixed value $W_{max} = 7.40 \times 10^5$ A/m for all steps, we find from Eq. (20) that $\mu_0 H_p = 0.76$ T. We continue this multistep process to obtain $\mu_0 H_p = 0.98, 0.81, 0.95, 0.89, 0.90$ and $0.894$ T after the third, fourth, … and eight steps respectively. We see that the last value, namely $\mu_0 H_p = 0.894$ T, differs by 0.5% and 0.7% only from its two previous values. Secondly, we see that the value of $H_p$ oscillates about $\mu_0 H_p = 0.894$ T so that the amplitude of oscillations move towards this value ($\mu_0 H_p = 0.894$) with increasing number of steps. The reason for the oscillations of $H_p$ lies in the fact that in a significant range of the values of $M^+(H)$ and $M^-(H)$ about $\mu_0 H_p = 0.894$ T $\Delta M(H)$ is a decreasing function of $H$. In fact, $M^-(H)$ contains a dip at $\mu_0 H = \mu_0 H_{min} = 0.35$ T (c. f., Fig. 1) but this is quite away from $\mu_0 H_p = 0.894$ T.

The above shows that for a Kim-like hysteresis loop it is rather an easy task to obtain the value of the full penetration field from Eq. (20). There may, however, be a trouble for example for a hysteresis loop of the type of fishtail behavior [9, 15]. In the present case, where we are restricted to the Kim model only, we shall follow the above multistep method with an accuracy of 99%. In this sense $\mu_0 H_p = 0.894$ T is the sought-for numerical solution for the hysteresis loop of Fig. 1.

The method of finding $H_p$ from Eq. (20) is useful for all values of $H_0$. On the other hand, in Ref. [10] we have obtained $H_p$ for $H_0 = 0$ only, which is

$$H_p = H_{min}/0.2912 \qquad \text{(for } H_0 = 0 \text{ only)} \qquad (21)$$

When we apply this equation to the hysteresis loop of Fig. 1 we find that $\mu_0 H_p = 0.35/0.2912 = 1.2$ T. This is a quite high value in comparison to $\mu_0 H_p = 0.894$ T found above. This means that the illustrative hysteresis loop of Fig. 1 does not correspond to $H_0 = 0$. Let us see whether it is indeed so or not. For this purpose we first of all calculate the value of $H_{0,hys}$ by using Eq. (8) with $\mu_0 H_p = 0.894$ T, $W_{max} = 7.40 \times 10^5$ A/m and $W_p(H = H_p) = 3.55 \times 10^5$ A/m. We find that $\mu_0 H_{0,hys} = 0.824$ T. Thereafter using the bridge equation, Eq. (12), we find $\mu_0 H_0 = 0.447$ T. This value of $H_0$ is not only different from zero, but is also comparable with $H_p$. Thus the hysteresis loop of Fig. 1 corresponds to a significantly larger value of $H_0$ than $H_0 = 0$. The validity of Eq. (21) is, therefore, not expected for the hysteresis loop of Fig. 1.

It may be noted that Eq. (20) is an important step for making the Kim model practically useful. Once we have found out $H_p$, we can easily calculate $H_{0,hys}$ (Eq. 8), $k_{hys}$ (Eq. 9), $J_{c,hys}(H)$ (Eq. 11), $H_0$ (Eq. 12) and $k$ (Eq. 2). If, on the other hand we do not have a hysteresis loop, and the values of $H_p$, $H_0$ and $k$ are known from some other method [6], then we can estimate $H_{0,hys}$ from Eq. (12) and $k_{hys}$ from Eq. (16).

2.4. Sample-width dependence of $H_p$

Although the dependence of the full penetration field on the width of a superconductor, Eq. (2), is well known for a long time, a lack of the knowledge of the values of the Kim constants $k$ and $H_0$ has remained a major hurdle in its wider use [3]. Now that we have worked out a method for calculating the Kim constants in a practically useful way, we consider the dependence of $H_p$ on $2a$ as given by Eqs. (2) and (16). For specificity, we consider Eq. (16), and rewrite it as

$$H_p = q_{hys}\sqrt{2k_{hys}a} = P_{hys}\sqrt{2a} \qquad (22)$$

where

$$q_{hys} = \sqrt{1+x^2} - x \qquad (23)$$

with

$$x = H_{0,hys}/\sqrt{2k_{hys}a} \qquad (24)$$

In order to realize the importance of the two factors of Eq. (22), $q_{hys}$ and $\sqrt{2k_{hys}a}$, or $P_{hys}$ and $\sqrt{2a}$, we first of all notice that, in general, both $H_{0,hys}$ and $k_{hys}$ depend upon the sample width [6]. In fact $H_{0,hys}$ and $k_{hys}$ are constants with respect to the magnetic field only [2, 6]. In Eq. (22)

the first factor $q_{hys}$ combines the sample-width dependence of $H_{0,hys}$ with the partial sample-width dependence of $k_{hys}$ such that the other part of the sample-width dependence of $k_{hys}$ lies in the second factor $\sqrt{2k_{hys}a}$. The importance of the factor $q_{hys}$ lies in the fact that it has a weak dependence on the sample width in comparison to that of $\sqrt{2k_{hys}a}$. For example, when $2a$ varies from 0 to ∞, then $q_{hys}$ varies from 0 to 1 only, as against the 0 to ∞ variation of $\sqrt{2k_{hys}a}$. Furthermore, the variations of $q_{hys}$ and $k_{hys}$ on $2a$ are opposite to each other in that while $q_{hys}$ increases with $2a$, $k_{hys}$ decreases (see Table 4 below). Thus, the product $P_{hys} = q_{hys}\sqrt{k_{hys}}$, which involves the complete sample-width dependence of both $H_{0,hys}$ and $k_{hys}$, will have even a weaker dependence upon the sample width than that of $q_{hys}$. Because of this, the main dependence of $H_p$ on the sample width will be given by $\sqrt{2a}$. This may be considered as the analog of the Bean result $H_p \propto 2a$.

## 3. Results and discussion

In order to clarify the importance of the hysteretic Kim model, as developed in the previous section, we have made numerical calculations for the $YBa_2Cu_3O_7$, $Bi_2Sr_2CaCu_2O_{8+\delta}$ and $Ba_{0.72}K_{0.28}Fe_2As_2$ superconductors on the basis of their hysteresis loops given respectively in Refs. [15], [14] and [16]. Here it may be noted that for the $Ba_{0.72}K_{0.28}Fe_2As_2$ superconductor the value of $H_{min}$ is zero. However, since $H_{min}$ itself does not enter in the expression of $H_{0,hys}$ (Eq. 8), there is no problem in applying the hysteretic Kim model to the $Ba_{0.72}K_{0.28}Fe_2As_2$ superconductor. In fact, the problem will be, for example with Eq. (21). But this equation is valid for $H_0 = 0$ only, whereas for the $Ba_{0.72}K_{0.28}Fe_2As_2$ superconductor we have found that $\mu_0H_0 = 1.36$ T (see below).

In table 1 we present the values of $W_{max}$, $\Delta M(0)$, $\Delta M(H_p)$ and $\mu_0 H_p$ for the considered superconductors. The values of $H_p$ are obtained in accordance with the illustration given in the previous section. In Table 2 we present values of $H_{0,hys}$ (Eq. 8), $k_{hys}a$ (Eq. 16) and $k_{hys}$ (Eq. 9) for the superconductors of Table 1. The values of the corresponding parameters in the original Kim model are given in Table 3. A comparison of the Tables 2 and 3 shows that the Kim constants have lower values in the original Kim model than those in the hysteretic Kim model. This is due to the bridge equation (Eq. 12), and due to Eq. (16).

3.1. Full penetration field

We have calculated values of $q_{hys}$ for the $YBa_2Cu_3O_7$, $Bi_2Sr_2CaCu_2O_{8+\delta}$ and $Ba_{0.72}K_{0.28}Fe_2As_2$ superconductors and presented them in Table 2. It is clear from these values that $q_{hys}$ varies at most by ~ 20% among the considered superconductors. On the other hand, the factor $\sqrt{2k_{hys}a}$ varies by ~ 500%. This means that the main dependence of $H_p$ on the sample width of the considered superconductors is governed by $\sqrt{2k_{hys}a}$.

According to Table 3, where we have presented values of

$$q = \sqrt{1 + (H_0^2/2ka)} - H_0/\sqrt{2ka}, \qquad (25)$$

which is the analog of $q_{hys}$, we expect the Kim model also to correspond mainly to the $H_p \approx \sqrt{2ka}$ relation with $q \approx 0.55$ to $0.75$.

In order to further illustrate the importance of Eq. (22) we consider the $Bi_2Sr_2CaCu_2O_{8+\delta}$ superconductor of Noetzel et al [14] for which we have found that $\mu_0 H_p = \mu_0 H_{p1} = 1.97$ T at $T =$ 4.2 K (c. f. Table 1). We compare this superconductor with the 90×90 μm$^2$ sample of the Bi$_2$Sr$_2$CaCu$_2$O$_{8+\delta}$ superconductor of Wang et al [3]. From an extrapolation of the $H_p$ versus $T$ data of Fig. 6 of Ref. [3] we find that $\mu_0 H_p = \mu_0 H_{p2} = 0.53$ T at $T = 4.2$ K for this sample. The sample width for the Bi$_2$Sr$_2$CaCu$_2$O$_{8+\delta}$ superconductor of Noetzel et al is $2a = 2a_1 = 1280$ μm (see Table 1), while that for the sample of Wang et al is $2a = 2a_2 = 90$ μm. From these values of $H_{p1}$, $H_{p2}$, $2a_1$ and $2a_2$ we find that $H_{p1}/H_{p2} = 1.97/0.53 = 3.72$, $2a_1/2a_2 = 1280/90 = 14.22$ and $\sqrt{2a_1/2a_2} = 3.77$. These ratios show that the two samples [3, 14] of the Bi$_2$Sr$_2$CaCu$_2$O$_{8+\delta}$ superconductor follow $H_p \approx \sqrt{2a}$, while the Bean relation (Eq. 13a) $H_p \propto 2a$ is too bad.

In order to see how the two samples of the Bi$_2$Sr$_2$CaCu$_2$O$_{8+\delta}$ superconductor having very large difference in their widths (1280 μm and 90 μm) follow the relation $H_p \approx \sqrt{2a}$ we proceed as follows. From Table 2 the values of $k_{hys}$ and $q_{hys}$ for the Bi$_2$Sr$_2$CaCu$_2$O$_{8+\delta}$ superconductor of Noetzel et al are $k_{hys,1} = 7.41 \times 10^{15}$ A$^2$/m$^3$ and $q_{hys,1} = 0.51$. In order to obtain the values of $k_{hys}$ and $q_{hys}$ for the 90×90 μm$^2$ sample of the Bi$_2$Sr$_2$CaCu$_2$O$_{8+\delta}$ superconductor of Wang et al at 4.2 K we use the bridge equation (Eq. 12) for obtaining $H_{0,hys}$ for $H_0 = 0$. (It may be recalled that for the Bi$_2$Sr$_2$CaCu$_2$O$_{8+\delta}$ superconductor of Wang et al $H_0 = 0$.) We get $\mu_0 H_{0,hys} = 0.3424 \times \mu_0 H_p = 0.3424 \times 0.53 = 0.1815$ T. Then using the values of $H_{0,hys}$ and $H_p$ in Eq. (16) we obtain $k_{hys} = k_{hys,2} = 3.36 \times 10^{15}$ A$^2$/m$^3$. Using these values of $H_{0,hys}$ and $k_{hys,2}$ in Eq. (23) we finally find $q_{hys} = q_{hys,2} = 0.77$ for the 90×90 μm$^2$ sample of the Bi$_2$Sr$_2$CaCu$_2$O$_{8+\delta}$ superconductor of Wang et al at 4.2 K. (Notice that in the considered case $q = 1.0$ because $H_0 = 0$.)

Now, from the values of $k_{hys,1}$, $k_{hys,2}$, $q_{hys,1}$ and $q_{hys,2}$ we find that the product $P_{hys}$ (c. f. Eq. 22) is $P_{hys,1} = 4.39 \times 10^7$ A/m$^{3/2}$ and $P_{hys,2} = 4.46 \times 10^7$ A/m$^{3/2}$ for the Bi$_2$Sr$_2$CaCu$_2$O$_{8+\delta}$ superconductors of Refs. [14] and [3] respectively. This means that the values of $P_{hys}$ $=q_{hys}\sqrt{k_{hys}}$ are almost the same in the considered two samples of the $Bi_2Sr_2CaCu_2O_{8+\delta}$ superconductor. In this sense it is the almost independence of $P_{hys}$ with respect to the sample width that has driven the relation $H_p \approx \sqrt{2a}$ in the two samples of the considered $Bi_2Sr_2CaCu_2O_{8+\delta}$ superconductor. Notice that the values of $q_{hys}$, 0.51 and 0.77, are not so independent of the sample width.

We now turn to the full penetration field of the $YBa_2Cu_3O_7$ superconductor. According to Table 1 we have found $\mu_0H_p = 0.57$ T for the $YBa_2Cu_3O_7$ superconductor of Senoussi et al [15]. On the other hand, these authors have estimated $\mu_0H_p$ to be about 0.15 T. The main reason for this large difference in the values of $H_p$ lies in the fact that these authors have used the Bean model for estimating the full penetration field. But the Bean model requires $H_{0,hys} \gg H_{irr}$ where $H_{irr}$ is the irreversibility field. Here we have compared $H_{0,hys}$ with $H_{irr}$ so that Eq. (1) remains independent of the magnetic field at least up to $H = H_{irr}$. In a strict sense we should have $H_{0,hys} \rightarrow \infty$ for the Kim model to approach the Bean model [13].

Turning to the present $YBa_2Cu_3O_7$ superconductor, Table 2 shows that $\mu_0H_{0,hys} = 0.54$ T for this superconductor. This is certainly much less than $H_{irr}$ of the $YBa_2Cu_3O_7$ superconductor [18]. Under such circumstances we can make only a rough estimate of $H_p$ by taking $J_{c,Bean}$ equal to that for the $H_{0,hys} \rightarrow \infty$ limit of Eq. (1). That is, by taking $J_{c,Bean} = k_{hys}/H_{0,hys} = J_{c,hys}(H=0)$. Using the value of $J_{c,hys}(H=0)$ from Table 2 we find that $J_{c,Bean} = 9.95 \times 10^{10}$ A/m$^2$. For the sample width of the $YBa_2Cu_3O_7$ superconductor of Table 1 this corresponds to $\mu_0 J_{c,Bean} a = 0.87$ T. Now following Senoussi et al [15] we estimate $H_p$ from $\mu_0 H_p \approx \mu_0 J_{c,Bean} a (1-N_c)$ where the demagnetization factor for the applied magnetic field parallel to the c direction, $N_c$ is 0.6 [15]. This gives $\mu_0 H_p \approx 0.35$ T, which is considerably larger than $\mu_0 H_p \approx 0.15$ T obtained by Senoussi et al [15].

3.2. Hysteretic Kim constant $H_{0,hys}$

An important quantitative feature of the constant $H_{0,hys}$ is that it will always be nonzero, as is very much clear from the above description, in particular, from Eq. (12). This has important consequences for the nonuniformity of the local internal field, which, in the hysteretic Kim model may be written as

$$p_{hys} = \sqrt{2k_{hys}a}/H_{0,hys} \qquad (26)$$

In fact, $p_{hys}$ is the analog of the nonuniformity parameter of the local internal field in the Kim model. That is $p_{hys}$ is the analog of [13]

$$p = \sqrt{2ka}/H_0 \qquad (27).$$

The importance of $p_{hys}$ may be realized by considering the data of the $Bi_2Sr_2CaCu_2O_{8+\delta}$ superconductor of Wang et al [3]. As we have seen above the value of $H_0$ is zero for this superconductor. Then from Eq. (27) we get $p = \infty$, which is certainly unrealistic. This is because if the local internal field has an infinite nonuniformity in some region, then the critical current density will be infinite in that region. But, in a superconductor the critical current density has always a finite upper bound, namely the depairing current density.

Let us see how Eq. (26) solves this problem. For specificity, we consider $H_p$ measured by Wang et al [3] for the 90×90 μm² sample of the $Bi_2Sr_2CaCu_2O_{8+\delta}$ superconductor at the temperature $T$=8 K. As we have mentioned above, the values of $\mu_0 H_{0,hys}$ and $k_{hys}$ for this situation are $\mu_0 H_{0,hys}$ = 0.103 T and $k_{hys}$ = 1.08X10$^{15}$ A²/m³. Using these values in Eq. (26) we obtain $p_{hys}$ = 3.78, which is not only finite but appears to be quite reasonable. In this way we have found a solution of the infinite nonuniformity of the local internal field.

The fact that $H_{0,hys}$ is always greater than zero is also important in providing a reasonable understanding of the mean bundle size of flux lines [19, 20]. In the original Kim model the mean bundle size is given by [21]

$$d = \sqrt{\Phi_0/\mu_0 H_0} \qquad (28)$$

The analog of this bundle size in the hysteretic Kim model may be written as

$$d_{hys} = \sqrt{\Phi_0/\mu_0 H_{0,hys}} \qquad (29)$$

Here also the situation is similar to that we faced in the case of the nonuniformity of the local internal field. More clearly, $d \rightarrow \infty$ for $H_0 = 0$, which is certainly unrealistic as the bundle size will not only be finite but also less than the penetration depth [1, 4, 21]. Now we consider the data of Wang et al [3] for the 90×90 μm² sample of the $Bi_2Sr_2CaCu_2O_{8+\delta}$ superconductor at the temperature $T$=8 K. As the value of $\mu_0 H_{0,hys}$ has been found above to be $\mu_0 H_{0,hys}$ = 0.103 T, Eq. (29) provides $d_{hys}$ = 142 nm as the value of the mean bundle size in the hysteretic Kim model. This is a reasonable value because for the $Bi_2Sr_2CaCu_2O_{8+\delta}$ superconductor the value of the penetration depth is ≈ 200 nm [22, 23]. This is yet another illustration of the importance of the hysteretic Kim model.

In Table 2 we have given the values of $p_{hys}$ and $d_{hys}$ for the $YBa_2Cu_3O_7$, $Bi_2Sr_2CaCu_2O_{8+\delta}$ and $Ba_{0.72}K_{0.28}Fe_2As_2$ superconductors of Table 1. The values of $p$ and $d$ are also given for these superconductors in Table 3. Although, from a quantitative viewpoint, both sets of values, ($p_{hys}$, $d_{hys}$) and ($p$, $d$), appear to be reasonable for the considered systems, the set ($p_{hys}$, $d_{hys}$) is based upon a consistent model.

### 3.3. Intergranular matrix

Up to now we have applied the hysteretic Kim model to the superconducting grains [15], and to superconducting single crystals [14, 16]. Many authors have employed the Kim model for the intergranular region also [5-8]. In order to show the importance of the hysteretic Kim model for the intergranular region we consider the low-field AC susceptibility data of Chen et al [6]. These authors have made a study of the AC susceptibility of the interganular region of four pieces of various sample widths (see Table 4) of the $YBa_2Cu_3O_{7-\delta}$ superconductor. These pieces are obtained from the same initial pellet. Let $P_1$, $P_2$, $P_3$ and $P_4$ denote these pieces with sample widths as $2a$ = 0.40, 1.29, 1.84 and 2.50 mm, respectively. The value of $2b$ is 2.60 mm for all the pieces.

We have obtained values of the hysteretic Kim constants $H_{0,hys}$ and $k_{hys}$ for all the sample pieces by taking the values of $H_0$ and $k$ from Chen et al [6], and present them in Table 4. From the values of $H_{0,hys}$ and $k_{hys}$ we have estimated the values of the intergranular critical current density $J_{c,hys}$ at $H = 0$ from Eq. (11). These values are given in Table 4. For comparison we have also given the values of $J_{c,Kim}(H = 0)$ for the intergranular matrix in Table 4. It is clear from Table 4 that $J_{c,hys}(H = 0)$ suffers at most a 10% variation among the considered four pieces of the $YBa_2Cu_3O_{7-\delta}$ superconductor. On the other hand, $J_{c,Kim}(H = 0)$ shows a very large variation of about 2000% among these pieces. Since the sample pieces are cut from the same initial pellet, all of them should correspond to similar values of the intergranular critical current density. While the hysteretic Kim model appears to support this trend very well, the Kim model does not. This shows that the hysteretic Kim model describes the intergranular matrix of the $YBa_2Cu_3O_{7-\delta}$ superconductor in a relatively much better way.

In Table 4 we have also given the values of $q_{hys}$ and $P_{hys} = q_{hys}\sqrt{k_{hys}}$. First of all we notice from these values that $q_{hys}$ and $k_{hys}$ vary oppositely with the sample width, as we have mentioned earlier also. Thus while $q_{hys}$ increases by a factor of 2.6 from sample $P_1$ to sample $P_4$, $P_{hys}$ increases by a factor of about 1.7 only. This shows that the combined term $P_{hys} = q_{hys}\sqrt{k_{hys}}$ causes the sample-width dependence of the full penetration field to tend towards the relation $H_p \propto \sqrt{2a}$.

## 4. Conclusions

In an earlier article [10] we have obtained expressions for the Kim constants on the basis of the hysteresis loop. In this article, we have found, by taking an example of the $H_0 = 0$ loop, that the method of Ref. [10] leads to a nonzero value of $H_0$. Such a difference arises in the values of $H_0$ for the same hysteresis loop because in Ref. [10] $H_0$ is governed by the condition that the critical current density should be physically reasonable down to $H = 0$. Corresponding to the two different values of $H_0$ for the same hysteresis loop we have considered two different versions of the Kim model. The first is the original Kim model, and the second is the hysteretic Kim model. While the original Kim model is expressed in terms of $H_0$, $k$ and $J_{c,Kim}(H)$, the hysteretic Kim model is expressed in terms of $H_{0,hys}$ (Eq. 8), $k_{hys}$ (Eq. 9) and $J_{c,hys}(H)$ (Eq. 11). The main point here is that the hysteretic Kim model is a practically useful model because of easily calculable expressions of $H_{0,hys}$ (Eq. 8), $k_{hys}$ (Eq. 9) and $H_p$ (Eq. 20). The importance of the hysteretic Kim model goes even further because we can also estimate the Kim constants $H_0$ and $k$ by using the bridge equation (Eq. 12).

The bridge equation is an important contribution of this article. In fact the bridge equation allows us to go from the Kim model to the hysteretic Kim model and vice versa. We have shown the importance of the bridge equation several times in the previous sections. Every time we have found that the bridge equation provides a reasonable insight in a given situation.

With readily available formulae for $H_{0,hys}$ (Eq. 8), $k_{hys}$ (Eq. 9), $H_p$ (Eq. 20), $H_0$ (Eq. 12) and $k$ (Eq. 2) we have made it practically interesting to study the dependence of the full penetration field on the sample width. In this connection the important point to note is that $H_{0,hys}$ and $k_{hys}$ also have dependence on the sample width. Same is true for $H_0$ and $k$ as is clear from Table 1 of Chen et al [6]. Absorbing the sample-width dependence of $H_{0,hys}$ and $k_{hys}$ in the factor $P_{hys}$ (Eq. 22) we have clarified that this factor ($P_{hys}$) depends only weakly on the sample width so that the main dependence of the full penetration field on the sample width is governed by the factor $\sqrt{2a}$.

Table 1: Values of the measurement temperature $T$, sample cross section $2a \times 2b$, maximum difference of hysteretic magnetization $W_{max}$, vertical $M$-$H$ loop widths $\Delta M(0)$ and $\Delta M(H_p)$, and the full penetration field $H_p$ for the $YBa_2Cu_3O_7$ [15], $Bi_2Sr_2CaCu_2O_{8+\delta}$ [14] and $Ba_{0.72}K_{0.28}Fe_2As_2$ [16] superconductors. The values of the sample dimensions $2a$ and $2b$ of the $Bi_2Sr_2CaCu_2O_{8+\delta}$ superconductor are obtained on the basis of Fig. 2 of Noetzel et al [14] by assuming that $b=a$. In order to show the difference of the (hysteretic) Kim model with the Bean model we have included symbolical values of the Bean model also in the last row of the table.

| System | $T$ (K) | $2a \times 2b$ (mm$^2$) | $W_{max}$ (emu/cm$^3$) | $\Delta M(0)$ (emu/cm$^3$) | $\Delta M(H_p)$ (emu/cm$^3$) | $\mu_0 H_p$ (T) |
|---|---|---|---|---|---|---|
| $YBa_2Cu_3O_7$ | 4.2 | 0.014×0.014 | 464 | 315 | 226 | 0.57 |
| $Bi_2Sr_2CaCu_2O_{8+\delta}$ | 4.2 | 1.28×1.28 | 1422 | 1297 | 832 | 1.97 |
| $Ba_{0.72}K_{0.28}Fe_2As_2$ | 8.0 | 2.85×4.20 | 3622 | 3622 | 1619 | 3.60 |
| Bean model | $T$ | $2a \times 2b$ | $\Delta M(0)$ | $\Delta M(0)$ | $\Delta M(0)$ | $G\Delta M(0)$ |

Table 2: Values of $H_{0,hys}$ (Eq. 8), $k_{hys}a$ (Eq. 16), $k_{hys}$ (Eq. 9), $H_{0,hys}^2/2k_{hys}a$, $q_{hys}$ (Eq. 23), $p_{hys}$ (Eq. 26), $d_{hys}$ (Eq. 29) and $J_{c,hys}(H=0)$ (Eq. 11) for the superconductors of Table 1. The last row shows the results for the Bean model.

| System | $\mu_0 H_{0,hys}$ (T) | $2\mu_0^2 k_{hys}a$ (T$^2$) | $k_{hys}$ (A$^2$/m$^3$) | $\dfrac{H_{0,hys}^2}{2k_{hys}a}$ | $q_{hys}$ | $p_{hys}$ | $d_{hys}$ (nm) | $J_{c,hys}(H=0)$ (A/m$^2$) |
|---|---|---|---|---|---|---|---|---|
| YBa$_2$Cu$_3$O$_7$ | 0.54 | 0.94 | 4.30×10$^{16}$ | 0.23 | 0.58 | 1.79 | 61.9 | 9.95×10$^{10}$ |
| Bi$_2$Sr$_2$CaCu$_2$O$_{8+\delta}$ | 2.78 | 14.82 | 7.41×10$^{15}$ | 0.52 | 0.51 | 1.39 | 27.3 | 3.33×10$^9$ |
| Ba$_{0.72}$K$_{0.28}$Fe$_2$As$_2$ | 2.91 | 34.05 | 7.64×10$^{15}$ | 0.25 | 0.62 | 2.00 | 26.7 | 3.28×10$^9$ |
| Bean model | ∞ | ∞ | ∞ | ∞ | 0 | 0 | 0 | $k_{hys}/H_{0,hys}$ |

Table 3: Values of $H_0$ (Eq. 12), $ka$ (Eq. 2), $k$, $H_0^2/2ka$, $q$ (Eq. 25), $p$ (Eq. 27), $d$ (Eq. 28) and $J_{c,Kim}(H=0)$ (Eq. 1) for the superconductors of Table 1. The last row shows the results for the Bean model.

| System | $\mu_0 H_0$ (T) | $2\mu_0^2 ka$ (T$^2$) | $k$ (A$^2$/m$^3$) | $\dfrac{H_0^2}{2ka}$ | $q$ | $p$ | $d$ (nm) | $J_{c,Kim}(H=0)$ (A/m$^2$) |
|---|---|---|---|---|---|---|---|---|
| YBa$_2$Cu$_3$O$_7$ | 0.30 | 0.67 | 3.06×10$^{16}$ | 0.13 | 0.70 | 2.73 | 83.1 | 1.28×10$^{11}$ |
| Bi$_2$Sr$_2$CaCu$_2$O$_{8+\delta}$ | 2.01 | 11.80 | 5.90×10$^{15}$ | 0.34 | 0.57 | 1.71 | 32.1 | 3.67×10$^9$ |
| Ba$_{0.72}$K$_{0.28}$Fe$_2$As$_2$ | 1.36 | 22.75 | 5.10×10$^{15}$ | 0.08 | 0.75 | 3.51 | 39.0 | 4.69×10$^9$ |
| Bean model | $\infty$ | $\infty$ | $\infty$ | $\infty$ | 0 | 0 | 0 | $k/H_0$ |

Table 4: Values of $H_{0,hys}$ (bridge equation, Eq. 12), $k_{hys}$ (Eq. 16), $J_{c,hys}(H=0)$ (Eq. 11), $q_{hys}$ (Eq. 23) and $P_{hys}$ (Eq. 22) for the intergranular matrix of the $YBa_2Cu_3O_{7-\delta}$ superconductor of Chen et al [6] for sample pieces $P_1$, $P_2$, $P_3$ and $P_4$ of various values of the sample width $2a$. Notice that $2b=2.60$ mm for all the pieces of the superconductor. For comparison values of $J_{c,Kim}(H=0)$, taken from Ref. [6] are also given.

| System pieces | $2a$ (mm) | $H_{0,hys}$ (A/m) | $k_{hys}$ ($A^2/m^3$) | $J_{c,hys}(H=0)$ ($10^6$ $A/m^2$) | $J_{c,Kim}(H=0)$ ($10^6$ $A/m^2$) | $q_{hys}$ | $P_{hys}$ ($10^4$ $A/m^{3/2}$) |
|---|---|---|---|---|---|---|---|
| $P_1$ | 0.40 | 1170 | 22.03 | 1.9 | 1.9 | 0.55 | 1.64 |
| $P_2$ | 1.29 | 519 | 10.59 | 2.0 | 3.6 | 0.65 | 2.11 |
| $P_3$ | 1.84 | 583 | 10.76 | 1.8 | 3.9 | 0.67 | 2.20 |
| $P_4$ | 2.50 | 457 | 9.37 | 2.0 | 37.0 | 0.91 | 2.78 |

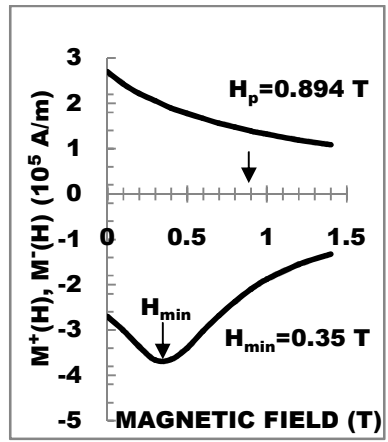

Fig. 1: A portion ($H=0$ to $H=2$ T) of a hysteresis loop having features of the Kim model (c. f., e. g., Fig. 6 of Chen and Goldfarb [13]). The values of $H_p$ and $H_{min}$ are indicated by arrows and given by numbers also.